# Multivariate Statistical Process Control Charts and the Problem of Interpretation: A Short Overview and Some Applications in Industry

S. Bersimis[1] J. Panaretos[2] and S. Psarakis[2]

**Abstract-** Woodall and Montgomery [35] in a discussion paper, state that multivariate process control is one of the most rapidly developing sections of statistical process control. Nowadays, in industry, there are many situations in which the simultaneous monitoring or control, of two or more related quality - process characteristics is necessary. Process monitoring problems in which several related variables are of interest are collectively known as Multivariate Statistical Process Control (MSPC).This article has three parts. In the first part, we discuss in brief the basic procedures for the implementation of multivariate statistical process control via control charting. In the second part we present the most useful procedures for interpreting the out-of-control variable when a control charting procedure gives an out-of-control signal in a multivariate process. Finally, in the third part, we present applications of multivariate statistical process control in the area of industrial process control, informatics, and business.

**Index Terms-** Quality Control; Process Control; Multivariate Statistical Process Control; Hotelling's T²; CUSUM; EWMA; PCA; PLS; Identification; Interpretation.

## I. INTRODUCTION

Most Statistical Process Control (SPC) approaches are based upon the control charting of a small number of variables, usually the final produce quality, and examining them one at a time. This is inappropriate for most process industry applications. It totally ignores the information collected on the process variables – possibly hundreds. The practitioner can not really study more than two or three charts to maintain process or product quality. It is very helpful that in practice, only a few events are driving a process at any one time; different combinations of these measurements are simply reflections of the same underlying events.

Multivariate SPC refers to a set of advanced techniques for the monitoring and control of the operating performance of batch and continuous processes. More specifically, multivariate SPC techniques reduce the information contained within all of the process variables down to two or three composite metrics through the application of statistical modeling. These composite metrics can then be easily monitored in real time in order to benchmark process performance and highlight potential problems, thereupon providing a framework for continuous improvements of the process operation.

Woodall and Montgomery [35] in a discussion paper, state that multivariate process control is one of the most rapidly developing sections of statistical process control. Harold Hotelling established multivariate process control techniques in his 1947 pioneering paper. Hotelling [11] applied multivariate process control methods in a bombsights problem.

Jackson [12] stated that any multivariate process control procedure should fulfill four conditions: a) an answer to the question: "Is the process in control?" must be available; b) an overall probability for the event "Procedure diagnoses an out-of-control state erroneously" must be specified; c) the relationships among the variables - attributes should be taken into account; d) an answer to the question: "If the process is out-of-control, what is the problem?" should be available. The Jackson's fourth condition is the most challenging problem at this time in the MSPC area, an appealing subject for many researchers in the last years, and the main topic under consideration in this article.

In Section 2, we discuss in brief the basic procedures for the implementation of multivariate statistical process control via control charting. In Section 3, we describe the most significant methods for the interpretation of an out-of-control signal. Furthermore, in Section 4, we present an extended set of application of multivariate statistical process control in the area of industrial process control. Finally, in Section 5 some concluding remarks are given with some points for further research.

## II. CONTROLLING AND MONITORING MULTIVARIATE PROCESSES USING CONTROL CHARTS

As we already stated, statistical process control techniques are widely used in industry. The most common process control technique is control charting. There are two distinct phases of control charting, Phase I and Phase II.

In Phase I, charts are used for retrospectively testing whether the process was in control when the first subgroups were being drawn. In this phase, the charts are used as aids to the practitioner, in bringing a process into a state of statistical in-control. Once this is accomplished, the control chart is used to define what is meant by statistical in-control.

---
[1]University of Piraeus, Department of Statistics and Insurance Science, Piraeus, Greece
[2]Athens University of Economic and Business, Department of Statistics, Athens, Greece

In Phase II, control charts are used for testing whether the process remains in control when future subgroups are drawn.

There are multivariate extensions for all kinds of univariate control charts (see e.g. Figure 1), such as multivariate Shewhart type control charts, multivariate CUSUM control charts, and multivariate EWMA control charts. In addition, there are unique procedures for the construction of multivariate control charts, based on multivariate statistical.

Shewhart type control charts for controlling the mean of an industrial process are usually based on the well known Mahalanobis distance statistic. The alternative forms of this statistic (distance) for the Phase I may be summarized as following: a) $\chi_i^2 = n(\overline{\mathbf{X}}_i - \boldsymbol{\mu})\boldsymbol{\Sigma}^{-1}(\overline{\mathbf{X}}_i - \boldsymbol{\mu})'$, for $i = 1,2,...,m$ rational subgroups, where $n$ is the sample size of each rational subgroup (with $n=1$ for individual observations), $\boldsymbol{\mu}$ is the vector of known means, $\boldsymbol{\Sigma}$ is the known covariance matrix and finally $\overline{\mathbf{X}}_i$ is the vector of samples means for the $i^{th}$ rational subgroup, b) $T_i^2 = n(\overline{\mathbf{X}}_i - \overline{\overline{\mathbf{X}}}_i)\overline{S}^{-1}(\overline{\mathbf{X}}_i - \overline{\overline{\mathbf{X}}}_i)'$, for $i = 1,2,...,m$, where $\overline{\overline{\mathbf{X}}}_i$ is the pooled vector of sample means calculated using the $n$ observed sample mean vectors, and $\overline{S}$ is the pooled sample covariance matrix. The $T_i^2$, and $\chi_i^2$ statistic represent the weighted distance of any point from the target (process mean under stable conditions). Under the assumption that the $m$ samples are independent and the joint distribution of the $p$ variables is the multivariate Normal, the $\chi_i^2$ follows a chi-square distribution with $p$ degrees of freedom and the $T_i^2$ follows $\frac{p(m-1)(n-1)}{mn-m-p+1}$ times an $F$ distribution with $p$, $mn-m-p+1$ degrees of freedom. Thus, the appropriate probability limits may be obtained using the known distributions of the corresponding statistic. In Figure 2, a control chart for a bivariate Normal process based on $T_i^2$ statistic is given.

Moreover, in the special case of a bivariate Normal process a control ellipse may be used. The ellipsoid presented in Figure 3, represents the 95% probability area of the bivariate Normal process.

Shewhart type control charts for controlling the variance of an industrial process are usually based either on the determinant of the covariance matrix $|S_i|$ which is called the generalized variance, or on the trace of the covariance matrix, $trS_i$, which is the sum of the variances of the variables.

For specific applications of these charts, as well as applications of other multivariate methods in quality improvement, the interested reader may consult Alt [1], Wierda [34], Lowry and Montgomery [15], Ryan [28], Bersimis [2], Bersimis et al. [3], Koutras et al. [14] or the more recent book by Mason and Young [17].

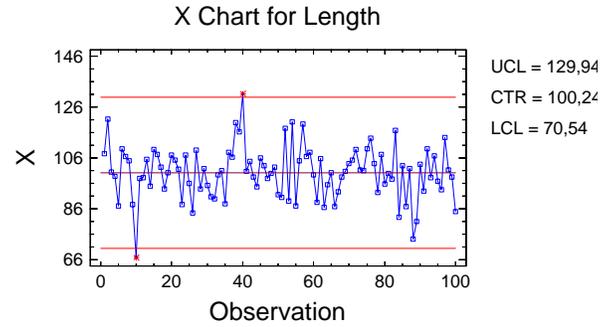

**Figure 1:** An univariate Shewhart Type Control Chart ($p=1$)

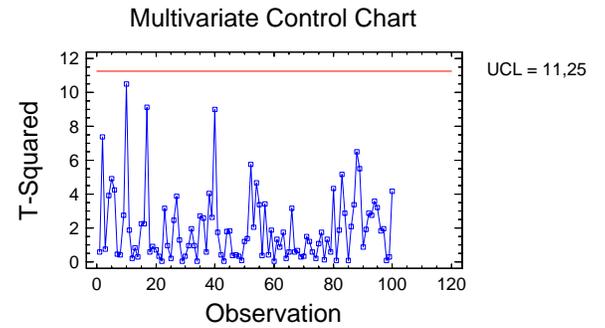

**Figure 2:** A multivariate Shewhart Type Control Chart ($p=2$)

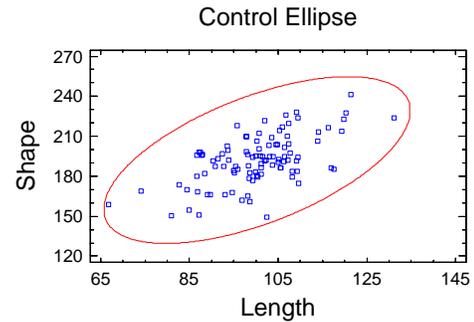

**Figure 3:** A Control Ellipse($p=2$)

Multivariate Shewhart type control charts use the information only from the current sample and they are relative insensitive to small and moderate shifts in the mean vector. Multivariate Cumulative Sum (MCUSUM) and Multivariate Exponentially Weighted Moving Average (MEWMA) control charts are developed to overcome this problem.

The multivariate CUSUM control charts are distinguished in two major categories. In the first case, the direction of the shift (or shifts) is considered to be known (direction specific schemes) whereas in the second the direction of the shift is considered to be unknown (directionally invariant schemes). Here we may note that the Shewhart type is always directionally invariant and the EWMA type control charts at the most of the cases.

Multivariate CUSUM schemes have been given by Woodall and Ncube [36] (the Multiple Univariate CUSUM Scheme, by Healy [10] (the CUSUM Based on the SPRT), by Crosier [5] (the CCV Scheme), as well as by Pignatiello and Runger [26] (the Mean Estimating CUSUM). The multivariate EWMA control chart proposed by Lowry et al. [16].

A problem with utilizing traditional multivariate Shewhart charts or multivariate CUSUM and EWMA schemes is that they may be impractical for high-dimensional systems with collinearities.

A common procedure for reducing the dimensionality of the variable space is the use of projection methods like Principal Components Analysis (PCA) and Partial Least Squares (PLS). These two methods are based on building a model from a historical data set, which is assumed to be in-control. After the model is built, the future observation is checked to see whether it fits well in the model. These multivariate methods have the advantage that they can handle process variables and product quality variables. Techniques such as PCA and PLS are used primary in the area of chemometrics (see eg Kourti[13], Wasterhuis et. al [33]) but they seem to be very promising in any kind of multivariate process.

### III. IDENTIFYING THE OUT-OF-CONTROL VARIABLE

In case that a univariate control chart gives an out-of-control signal, the practitioner may easily conclude what the problem is and give a solution since a univariate chart is related to a single variable. In a multivariate control chart the solution to this specific problem is not straightforward since any chart is related to a number, greater than one, of variables and also correlations exist among them. In this section we present methods for detecting, which of the $p$ variables is out of control.

A first approach to this problem was proposed by Alt [1] who suggested the use of Bonferroni limits. Hayter and Tsui [9] extended the idea of Bonferroni-type control limits by giving a procedure for exact simultaneous control intervals for each of the variable means, using simulation. A similar control chart is the Simulated MiniMax control chart presented by Sepuldveda and Nachlas [29].

Alt [1] and Jackson [12] discussed the use of an elliptical control region. However, this process has the disadvantage that it can be applied only in the special case of two quality characteristics. An extension of the elliptical control region as a solution to the interpretation problem is given by Chua and Montgomery [4].

Today the use of $T^2$ decomposition proposed by Mason et al. [18] is considered as the most valuable. The main idea of this method is to decompose the $T^2$ statistic into independent parts, each of which reflects the contribution of an individual variable.

The problem with this method is that the decomposition of the $T^2$ statistic into $p$ independent $T^2$ components is not unique. Thus, Mason et al. [19] give an appropriate computing scheme that can greatly reduce the computational effort. Mason et al. [20] presented an alternative control procedure for monitoring a step process, which is based on a double decomposition of Hotelling's $T^2$ statistic. Mason and Young [21] showed that by improving the model specification at the time that the historical data set is constructed, it may be possible to increase the sensitivity of the $T^2$ statistic to signal detection. The methodologies of Murphy [24], Doganaksoy et al. [6], Timm [31] and Runger et al. [27], are special cases of Mason's et al. [18] partitioning of $T^2$.

Jackson [12] proposed the use of principal components for monitoring a multivariate process. Since the principal components are uncorrelated, they may provide some insight into the nature of the out of control condition and then lead to the examination of particular original observations. Tracy et al. [32] expanded the previous work and provided an interesting bivariate setting in which the principal components have meaningful interpretations.

Principal components can be used to investigate which of the $p$ variables are responsible for an out-of-control signal. Until nowadays, writers have proposed various methods which use principal components for interpreting an out-of-control signal. The most common practice is to use the first $k$ most significant principal components, in the case that a $T^2$ control charts gives an out-of-control signal. The principal components control charts, which were analyzed in the corresponding section, can be used. The basic idea is that the first $k$ principal components can be physically interpreted, and named. Therefore, if the $T^2$ chart gives an out-of-control signal and for example the chart for the second principal component gives also an out-of-control signal, then from the interpretation of this component, a direction can be taken for which variables are the suspect to be out-of-control. The practice just mentioned transforms the variables into a set of attributes. The discovery of the assignable cause that led to the problem, with this method, demands a further knowledge of the process itself from the practitioner. The basic problem of this method is that the principal components have not always a physical interpretation.

According to Jackson [12], the procedure for monitoring a multivariate process using PCA can be summarized as follows: For each observation vector, obtain the $z$-scores of the principal components and from these compute $T^2$. If this is in control, continue processing. If it is out-of-control, examine the $z$-scores. As the principal components are uncorrelated, they may provide some insight into the nature of the out-of-control condition and may then lead to the examination of particular original observations.

Kourti and MacGregor [13], provide a different approach based on principal components analysis. The $T^2$ is expressed in terms of normalized principal components scores of the multinormal variables. When an out-of-control signal is received, the normalized score with high values are detected, and contribution plots are used to find the variables responsible for the signal. A contribution plot indicates how each variable involved in the calculation of that score contributes to it. Computing variable contributions eliminates much of the criticism that principal components lack of physical interpretation. This approach is particularly applicable to large ill conditioned data sets due to the use of principal components. Contribution plots are also explored by Wasterhuis et al. [33].

Maravelakis et al. [22] proposed a new method based on principal components analysis. Theoretical control limits were derived and a detailed investigation of the properties and the limitations of the new method were given. Furthermore, a graphical technique which can be applied in these limiting situations was provided.

Fuchs and Benjamini [7], presented a method for simultaneously controlling a process and interpreting a out-of-control signal. This is a new chart (graphical display) that emphasizes the need for fast interpretation of an out-of-control signal. The multivariate profile chart (MP chart) is a symbolic scatterplot. Summaries of data for individual variables are displayed by a symbol, and global information about the group is displayed by the location of the symbol on the scatterplot. A symbol is constructed for each group of observations. The symbol is an adoption of a profile plot that encodes visually the size and the sign of each variable from its reference value. Fuchs and Kenett [8], developed a Minitab macro for creating MP charts.

Sparks et al. [30], presented a method for monitoring multivariate process data based on the Gabriel biplot. They illustrated the use of the biplot on an example of industrial data. Nottingham et al. [25], developed radial plots as SAS-based data visualization tool that can improve process control practitioner's ability to monitor, analyze, and control a process. Finally, Maravelakis and Bersimis [23] presented an algorithm using the well known Andrews curves for solving the problem of interpreting an out-of-control signal.

## IV. Applications of Multivariate SPC Techniques in the Industrial Environment

In this Section, we will discuss in brief an application of the multivariate SPC techniques in industry. Specifically, we will analyze a three-variable real case relative to the quality of a chemical process.

In the beginning, we proceed with a Phase I analysis. In this Phase, our interest is to estimate the parameters (the mean vector and the covariance matrix), to check for the existence of dependence among the variables and finally, to check the validity of the assumption of multivariate normality.

As a first step in the analysis, we must test the assumption of multivariate normality of the three variables. If the three variables come from a 3-dimensional Normal distribution the $T_i^2$ values must follow a chi-square distribution with 3 degrees of freedom. As we may observe in Figure 4 the hypothesis that the $T_i^2$ values follow a chi-square distribution with 3 degrees of freedom can not be rejected. The second step in the analysis is the calculation of the correlation matrix. The application of a multivariate control chart is needed only if the three variables are strongly related. As we may easily see in Figure 5 the three variables are strongly related with correlation coefficients equal to $r(X_1, X_2) = 0.5929, r(X_1, X_3) = 0.7287, r(X_2, X_3) = 0.8170$. Since there is a strong correlation among the three variables, we must use a multivariate procedure for controlling the mean level of the process. Thus, we may apply a Shewhart type control chart and evaluate the state of the process. In Figure 6 we may see that the process till the 100th time point is in-control. Using the fact that the process is in-control we may use the estimated parameters in the Phase II analysis.

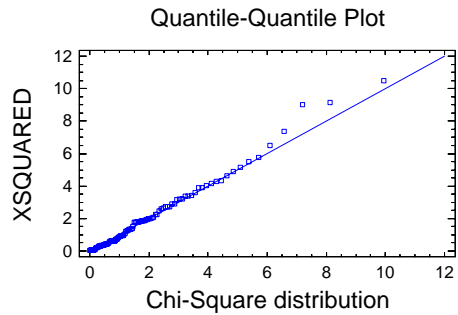

**Figure 4:** Quantile-Quantile Plot for the Values of $T_i^2$

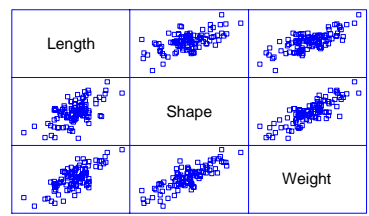

**Figure 5:** Matrix Scatter Plot for the three variables

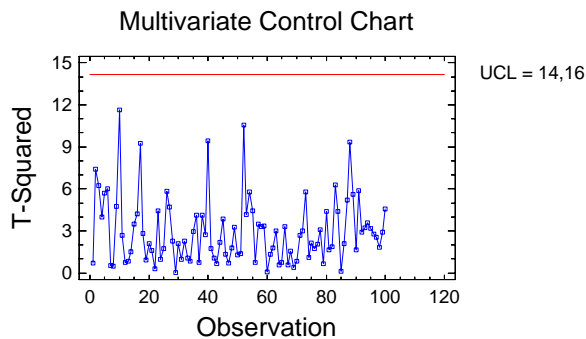

**Figure 6:** A multivariate Shewhart Type Control Chart (*p*=3)

In Phase II, the multivariate Shewhart type control chart is used for testing whether the process remains in control from the 101st time point and after. At the 101st point as we may easily observe in Figure 7 the process moved to an out-of-control state.

At this time point the practitioner must find the variable that contributed in the out-of-control signal.

As we presented in Section 3 there are too many options for identifying the variable responsible for the out-of-control message.

In this application, we will use the methods proposed by Maravelakis and Bersimis [23] and Maravelakis et al. [22]. Both of these procedures are classified to the graphical techniques for interpreting the out-of-control signal.

The ratio $F_{13}$, which connects the third variable and the first principal component, is charted in Figure 8. In Figure 9, the other two ratios $F_{12}$, $F_{11}$, are presented. From this figures it is clear that the third variable is responsible for the out-of-control signal at the 101$^{st}$ sampling point.

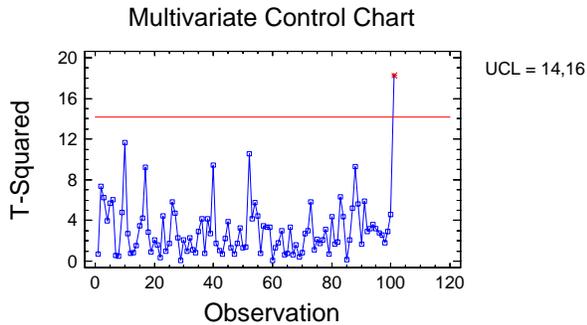

**Figure 7:** A multivariate Shewhart Type Control Chart ($p$=3)

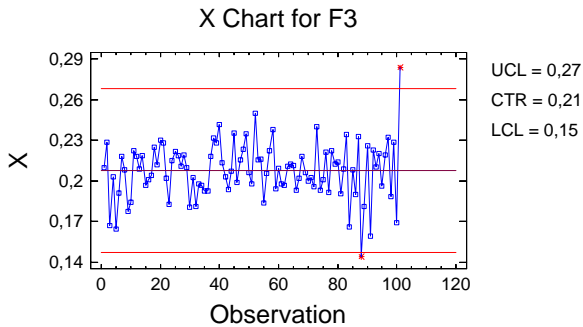

**Figure 8:** Identifying the out-of-control variable (Procedure based on [22])

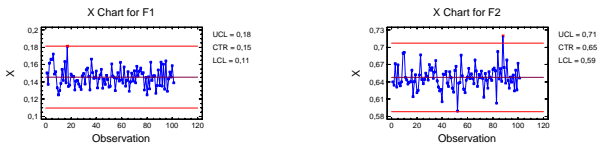

**Figure 9:** Identifying the out-of-control variable (Procedure based on [22])

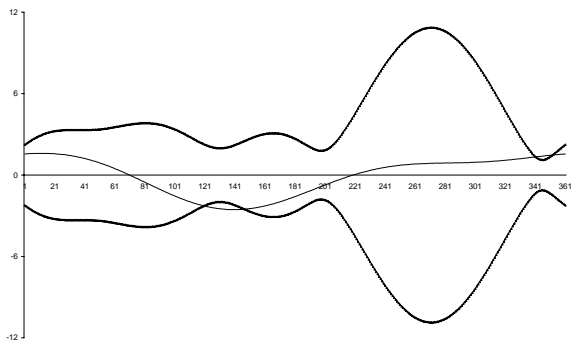

**Figure 10:** Identifying the out-of-control variable (Procedure based on [23])

In Figure 10, the graphical display of the procedure based on Maravelakis and Bersimis [23] is presented. According to this procedure each of the three variables corresponds to specific intervals of the $[-\pi, \pi]$. The interval from 210 to 250 corresponds to the third variable implying that the third variable is responsible for the out-of-control signal. In conclusion, the two methods gave us the same result, thus the practitioner has to check for possible assignable causes at the mechanisms related to the third variable.

## V. COMMENTS

Interesting areas for further research in the domain of multivariate SPC are robust design of control charts and nonparametric control charts. The research for multivariate attributes control charts is also a promising task. The problem of interpreting an out-of-control signal is an open area which needs further investigation.